\documentclass[twoside,fleqn]{ActaStyle}
\usepackage{times,cite}
\usepackage{lscape}             
\usepackage{graphicx,epsfig}    
\usepackage[tbtags]{amsmath}    

\def\authorheading{S.~Schadmand}
\def\shorttitle{WASA at COSY}
\begin{document}
\pagerange{1}{6}
\title{%
WASA at COSY}

\author{%
 S.~Schadmand\email{s.schadmand@fz-juelich.de}
 {\em ~~for the WASAatCOSY collaboration}
}
{
 Institut f\"ur Kernphysik, Forschungszentrum J\"ulich, Germany
}

\day{November 18, 2005}

\abstract{%
The Wide Angle Shower Apparatus (WASA) has been transferred from the storage ring
CELSIUS (TSL, Uppsala, Sweden) to the cooler synchrotron COSY (FZ J\"ulich, Germany).
In this presentation, the status and planning of the WASAatCOSY project are reported.
The main physics aspects are rare decays of $\eta$ and $\eta$' mesons
and isospin violation in deuteron-deuteron interactions.
}


\section{Introduction}

WASAatCOSY will provide unique scientific possibilities for
research in hadron physics with hadronic probes.
The COSY beam energy range, the phase
space cooling and the availability of (polarized) proton and
deuteron beams are essential aspects for high luminosity
measurements with low background.
The WASA detector provides nearly full solid angle coverage for both
charged and neutral particles which is necessary for kinematically
complete measurements of multiparticle final states.
Furthermore, the use of frozen pellets of liquid hydrogen and deuterium
as the target minimizes background from secondary reactions, at the same
time allowing high luminosity conditions.

The physics that will be investigated comprises
symmetries and symmetry breaking and hadron structure and interactions.
Finding and further investigating specific hadronic bound systems will,
together with corresponding progress in theory,
provide fundamental insight into how nature makes hadrons.
The physics proposal \cite{WASAprop} concentrates on the study of
symmetry breaking as the primary objective, in particular in $\eta$
and $\eta{}'$ decays and meson production.

The technical solution for the project WASAatCOSY
concerns the transfer of the WASA detector system and pellet target to COSY
and the installation at an internal target position.
After tests of the individual components, necessary repairs and
modifications, and preparations at COSY, the detector is foreseen to be
installed until end of July 2006.
Six weeks of commissioning beam time divided into three blocks
and an additional week of dedicated
machine development until the end of 2006 are foreseen
to be able to start the physics program at the beginning of 2007.

\section{Experimental Facility}

COSY is a cooler synchrotron and storage ring operated
at the Institute for Nuclear Physics (IKP)
of the Forschungszentrum J\"ulich.
The accelerator complex comprises an isochronous
cyclotron (JULIC), used as an injector, a race track shaped cooler
synchrotron with a circumference of 184~m, and
internal and external target stations~\cite{RMaier97}.
COSY delivers beams of polarized and unpolarized protons and deuterons
in the momentum range between 0.3~GeV/c and 3.7~GeV/c.
The ring can be filled with up to $10^{11}$~particles leading to
a typical luminosity of $10^{31}\,\mathrm{cm^{-2}s^{-1}}$
when using an internal cluster target.
Beams can be phase-space cooled by means of electron cooling
at injection energy as well as stochastic cooling
at high energies. Typical beam preparation times, including injection,
accumulation and acceleration, are of the order of a few
seconds, while the beam lifetime with a cluster target is between several
minutes and an hour.

\begin{figure}[htb]
\includegraphics[width=\textwidth,clip]{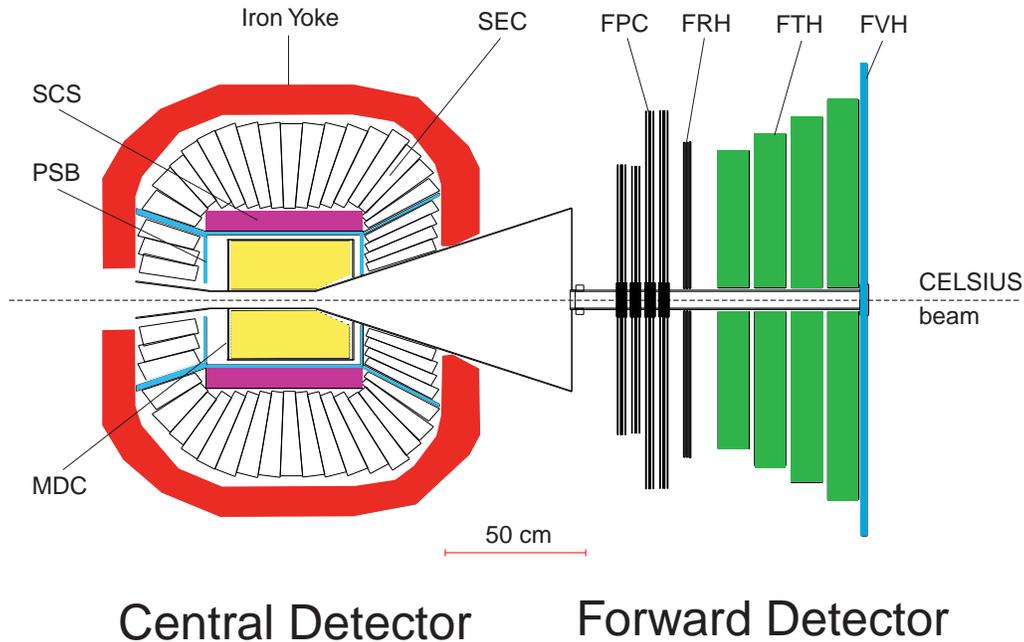}
\caption{
 Cross section of the WASA detector at CELSIUS.
 The central detector built around the interaction point (at the left)
 is surrounded by an iron yoke. The layers of the forward detector are
 visible on the right hand side.
 The individual components are described in \cite{WASAprop}.
}\label{fig:wasa}
\end{figure}

WASA at COSY will provide unique scientific possibilities for
research in hadron physics with hadronic probes.
It combines COSY, with proton and deuteron beams with energies sufficient
to cover the strange quark sector including $\phi$-mesons, and
WASA, a close to $4\pi$ detector for both photons and charged
particles, Fig.~\ref{fig:wasa}.
%

The WASA detector \cite{1} is being transferred from CELSIUS to COSY.
The actual transfer is ahead of schedule with the detector components
having arrived at COSY beginning of September 2005.
To start the physics program for WASA at COSY with the beginning of
2007, preparations both of the detector and at COSY have to be
finished by summer 2006, when the installation at COSY is foreseen
leaving the second half of 2006 for commissioning runs.
Preparations that do not require the actual detector hardware had already
been proceeding as planned. Modifications and repairs on detector elements are
presently being started.

The WASA detector will be installed in one of the straight sections of COSY
in front of the electron cooler at the former location of an RF cavity which has
already been removed. The location ensures a dispersion-free target position.
A rail system for moving the electromagnetic calorimeter during installation or
service will be installed in the COSY tunnel during the 2005/2006 winter shutdown.
Also, the corresponding rails outside of the  tunnel will be completed during that
time.
The support structure for the detector and the pellet target control platform used
at CELSIUS have to be modified to adapt to the concrete shielding of COSY.
With the modified support structure, the pellet generator will be 
decoupled mechanically, to avoid any influence from vibrations caused 
by vacuum pumps. 
The new support structure will be completed in March 2006.
End of May 2006 vacuum components will be available for installation.

The target will be set in operation as soon as possible
to ensure reliable pellet target performance after the transfer.
The reassembly of components will take place immediately after arrival
of the hardware.
A test setup is being built which can be operated until installation.
Minor changes of the target are planned in this period. The know-how for
the fabrication of the most delicate components, i.e. the nozzle and the vacuum
injection capillary, will be transferred.

The energy measurement in the WASA Forward Detector is based on the
dE/E technique using an arrangement of plastic scintillators. 
Track coordinates are determined by straw chambers. 
The thickness of the plastic scintillators was optimized for CELSIUS energies, i.e.
for protons with kinetic energies up to 500~MeV.
To adapt to the higher proton energies involved, for example,
using the tagging reaction $pp\to pp\eta^\prime$,
several modifications of the Forward Detector are presently under discussion.
     An upgrade of the scintillator arrangement by an additional layer
will increase the energy for stopping protons from 300 to 350~MeV and
improve the resolution for protons in the range of 300 to 400~MeV by 35\%.
The improvement at higher energies will be in the order of 25\% for protons
with a kinetic energy of 900~MeV with a resulting energy resolution
of $\sigma_E/E\approx$10\%.
    A Cerenkov detector with a sensitivity between 500~MeV and 1200~MeV
is considered with a resolution of 5\% at proton energies above 500~MeV.
A prototype detector has been built and parasitic in-beam tests are scheduled.
Both, the forward and central trackers will be equipped with new,
state-of-the-art electronics modules replacing the present preamplifier
and discriminator boards.

To accommodate for the higher trigger rates expected and in view of considerable
maintenance problems with the existing acquisition system, the data acquisition
will be replaced by a third generation data acquisition developed at J\"ulich.
Digitizing modules are being developed by the J\"ulich electronics department (ZEL),
the Department of Radiation Sciences (ISV), and The Svedberg Laboratory (TSL) 
in Uppsala. 
Prototypes will be available for tests with the
WASA detector components end of October 2005. After prototype iteration
being completed end of March 2006, mass production will start.
The readout hardware should be available at installation time.

\section{Commissioning}

During the commissioning phase for WASAatCOSY, the experimental facility
will be prepared for starting the physics program in 2007. All detector
components will be set into operation and the performance will be verified.
Although performance testing and improvement are the major
goals of the commissioning phase, physics results may already be extracted. 
Thus, the commissioning phase has been staged into four parts, 
each focusing on different performance aspects of the detector and
related to the different physics issues that are vital 
for the WASA physics program.

The operation of the WASA pellet target at COSY requires careful optimization
of beam conditions, including a thorough study of beam--target
interaction, especially in view of luminosities in the order
of $10^{32}$cm$^{-2}$s$^{-1}$
that are needed for the key parts of the experimental program.
For this, dedicated machine development is mandatory to ensure
high quality beam conditions even in the commissioning phase.
The machine development is scheduled immediately
after installation, to allow elementary tests like trigger and count rates.

During the first commissioning stage, the focus will be
on $\eta$ decays, primarily to compare the performance of the detector with the
experience from WASA operation at CELSIUS. For this purpose, the
same beam energy will be chosen as during data taking at CELSIUS
(T$_p$ = 1360~MeV). Calibration decay channels $\eta \to \gamma\gamma$
and $\eta \to 6 \gamma$ should be supplemented by a trigger focusing
on both neutral and charged $\eta \to 3\pi$ decays.
Thus, the detector performance in both neutral and charged particle
reconstruction can be tested and the quality achieved in terms of resolution
can be compared to results obtained at CELSIUS \cite{3}. One physics
issue involved is the determination of the slope parameter of the Dalitz
plot distribution for the neutral $3\pi$ decay. Preliminary data obtained at
the KLOE facility \cite{4} disagree with the Crystal Ball result \cite{5}.
Due to the lower statistics obtained at CELSIUS, it will not be possible to
discriminate between the two from existing WASA data.

At higher energies (T$_p$ = 2540~MeV),
the measurement of the dominant $\eta^\prime \to \eta 2\pi$ decay channels probes
both the performance of the modified
Forward Detector in view of $\eta^\prime$ tagging via missing mass, and the $\eta^\prime$
reconstruction from the decay products involving the Central Detector.
The decay $\eta^\prime \to \gamma\gamma$ decay will be measured in parallel 
to compare with results for the $\eta$ obtained during the second commissioning 
stage.
The physics motivation for studying the $\eta 2\pi$ decay is a determination of the
Dalitz plot parameters that can be compared with existing data \cite{6,7,8}.
Simultaneously, first data can be taken for the $\eta^\prime \to \pi^+\pi^-\gamma$
mode that is related to the manifestation of higher order anomalies \cite{WASAprop}.

A test of isospin violation in the reaction $\vec{d}d \to {}^4He \pi^o$
will be a further key experiment for WASA at COSY.
This measurement requires a deuterium
pellet target, high luminosity, and the ability to cleanly identify
$^4He$ recoils in coincidence with a neutral pion, and an effective background
suppression. While a series of experiments is necessary to address these
issues properly, $2\pi$ production is an ideal starting point to study most of
these aspects. In addition, the $2\pi$ data prepare for future measurements
of the reaction $dd \to {}^4He \pi\eta$ to study $a_o-f_o$ mixing.
As a surplus, aspects
of the ABC effect \cite{9,10,11} can be studied with the $\pi\pi$ system being produced
in a pure isospin zero state.
By choosing the appropriate beam
energy, the data can be directly compared to data taken during the last
period of data taking at CELSIUS \cite{12}.

\section{Summary}

The WASA detector has been transferred from CELSIUS to
COSY. After tests of the individual components, necessary repairs and
modifications, and preparations at COSY, the detector is foreseen to be
installed at COSY until end of July 2006.
The WASA detector at COSY is bound to produce physics results as soon as
possible. The time schedule until the start of the experimental program is
ambitious but feasible. 
Preparations for taking data are proceeding as planned.
In order to start the physics program in 2007, a commissioning phase
has been scheduled for the second half of 2006.



\begin{thebibliography}{99}
%
\bibitem{WASAprop}
Adam, H.-H., et al., Proposal for the Wide Angle Shower Apparatus (WASA)
at COSY--J\"ulich | 'WASA at COSY', COSY Proposal (2004), e-Print
Archive: nucl-ex/0411038.
\bibitem{RMaier97} R. Maier, Nucl. Instr. Meth.  A 390,  1  (1997).
%
\bibitem{1} Zabierowski, J., et al., Phys. Scripta, T99, 159-168 (2002).
\bibitem{3}  Jacewicz, M., Measurement of the Reaction $pp \to pp\pi^+\pi^-\pi^o$
 with CELSIUS/WASA at 1.36 GeV/c, PhD thesis, Uppsala University (2004).
\bibitem{4}
 Giovannella, S., et al., Kloe results on $f_0$(980), $a_0$(980) scalars and $\eta$ decays
 (2005), e-Print Archive: hep-ex/0505074.
\bibitem{5} Tippens, W. B., et al., Phys. Rev. Lett., 87, 192001 (2001).
\bibitem{6}  Kalbeisch, G. R., Phys. Rev., D 10, 916-920 (1974), and references therein.
\bibitem{7}  Alde, D., et al., Phys. Lett., B 177, 115-119 (1986).
\bibitem{8}  Briere, R. A., et al., Phys. Rev. Lett., 84, 26-30 (2000).
\bibitem{9}  Abashian, A., Booth, N. E., and Crowe, K., Phys. Rev. Lett., 5, 258-260 (1960).
\bibitem{10}  Abashian, A., et al., Phys. Rev., 132, 2296-2304 (1963).
\bibitem{11}  Booth, N. E., Phys. Rev., 132, 2305-2308 (1963).
\bibitem{12}  Th\"orngren-Engblom, P., Study of double pion production in dd and pd
reactions, CELSIUS Proposal CA 65 (2000).

\end{thebibliography}
\end{document}